\documentclass[12pt]{article}
\usepackage{amsfonts}
\usepackage{anysize}
\usepackage{amsmath,amssymb,amsbsy,amscd,amsthm,dsfont}
\usepackage{enumerate}
\usepackage{graphicx}
\usepackage{float}

\def\>#1{{\mathbf{#1}}}

\begin{document}
\title{On the tomographic description of classical fields}
\author{A. Ibort$^{a,\dagger}$, A. Lopez-Yela$^a$, V.I. Man'ko$^b$, G. Marmo$^{c,\ddagger}$, A. Simoni$^c$,\\
 E.C.G. Sudarshan$^d$, F.Ventriglia$^c$\\
{\footnotesize \textit{$^a$Departamento de Matem\`{a}ticas, Universidad
Carlos III de Madrid, }}\\
{\footnotesize \textit{Av.da de la Universidad 30, 28911 Legan\'{e}s,
Madrid, Spain }}\\
{\footnotesize {(e-mail: \texttt{albertoi@math.uc3m.es, alyela@math.uc3m.es})}}\\
{\footnotesize \textit{$^b$P.N.Lebedev Physical Institute, Leninskii
Prospect 53, Moscow 119991, Russia}}\\
{\footnotesize {(e-mail: \texttt{manko@na.infn.it})}}\\
\textsl{{\footnotesize {$^c$Dipartimento di Scienze Fisiche dell' Universit%
\`{a} ``Federico II" e Sezione INFN di Napoli,}}}\\
\textsl{{\footnotesize {Complesso Universitario di Monte S. Angelo, via
Cintia, 80126 Naples, Italy}}}\\
{\footnotesize {(e-mail: \texttt{marmo@na.infn.it, simoni@na.infn.it,
ventriglia@na.infn.it})}}\\
\textsl{{\footnotesize{Physics Department, Center for Particle Physics, University of Texas, Austin, Texas, 78712 USA.}}}\\
{\footnotesize{(e-mail: \texttt{bhamathig@gmail.com})}}\\
{\footnotesize{$^\dagger$ This work was partially supported by}}\\
 {\footnotesize{MEC grant MTM2010-21186-C02-02 and QUITEMAD programme.}}\\
{\footnotesize{$^\ddagger$ G.M. would like to acknowledge the support provided by}}\\ 
{\footnotesize{the Santander/UCIIIM Chair of Excellence programme 2011-2012.}}}
\maketitle





\begin{abstract}
After a general description of the tomographic picture for classical
systems, a tomographic description of free classical scalar fields is
proposed both in a finite cavity and the continuum. The tomographic
description is constructed in analogy with the classical tomographic picture
of an ensemble of harmonic oscillators. The tomograms of a number of
relevant states such as the canonical distribution, the classical
counterpart of quantum coherent states and a new family of so called
Gauss--Laguerre states, are discussed. Finally the Liouville equation for field
states is described in the tomographic picture offering an alternative
description of the dynamics of the system that can be extended naturally to
other fields.
\end{abstract}



\section{Introduction}

Recently it has been shown the equivalence between the tomographic picture
of quantum states and the various standard representations of them: Schr\"{o}%
dinger \cite{Sc26}, Heisenberg \cite{He27}, Wigner \cite{Wi32}, etc. (see
for instance \cite{Pedatom}, \cite{VentriPositive} and references therein).
In this paper we try to extend such description to classical fields. In
particular we will discuss the tomographic description of the real scalar
Klein--Gordon field inspired by the tomographic description of an ensemble
of harmonic oscillators. In fact classical and quantum field states are
usually considered as classical and quantum mechanics applied to describing
these states for systems with infinite number of degrees of freedom (the
field modes). Thus a state of a classical free field when restricted to
consider just a finite number of modes can be treated as a statistical
ensemble of harmonic oscillators.

This attempt will generalize the description of classical (or quantum)
states in two directions. On one side, describing classical field states
involves dealing with an infinite number of degrees of freedom and on the
other, a covariant treatment of fields implies taking into the description
of the state its dynamical evolution. In order to show how to proceed with
this task we will analyze the foundations of tomography for a classical
system with a finite number of degrees of freedom and we will extend
straightforwardly such construction to deal with classical fields.

The tomographic description of classical systems presented here will be
directly inspired by the Radon transform, so that our construction can also
be considered as an infinite dimensional extension of the Radon transform.
The transition to quantum fields should proceed using similar ideas in the
realm of quantum mechanical systems, however we will leave such analysis to
a subsequent paper. Classical and quantum standard descriptions of the
fields are dramatically different. The states of classical modes are
identified with probability densities and the states of quantum modes are
identified with Hermitian trace-class nonnegative density operators (or
density matrices). The observables in the tomographic picture of quantum
mechanics are tomographic symbols of corresponding operators which are
constructed by means of a specific star--product scheme \cite{JPA2002}. The
analogous tomographic representation of classical system states by means of
classical Radon transform \cite{Ra17} of the classical probability density $%
\rho(\omega)$ is also available \cite{Manko-Tombesi}, \cite{Me00}, \cite%
{Pedatom}.

Till now the classical and quantum field states have not been considered in
the tomographic probability representations, except in early attempts \cite%
{Ma98}, \cite{St05} and more recently when applying the quantum Radon
transform to study tomographic symbols of creation and annihilation field
operators for bosons and fermions \cite{An09} \cite{Ma09}. The aim of our
work is to extend the tomographic approach to the case of quantum and
classical systems with infinite number of degrees of freedom and to
introduce for classical and quantum fields the tomographic probability
density functionals determining their states. We will also find the
tomographic form of the classical field Liouville equation for the
tomographic probability density functionals.

The paper is organized as follows. In section 2 a generalized description of
the tomography of classical systems inspired on the Radon transform will be
described. In this picture the description of states of a physical system as
normalized positive functionals on the algebra of observables of the system
is paramount. Notice that such framework is common to both classical and
quantum systems. The tomographic description of a family of states, similar
to coherent states, for an ensemble of independent harmonic oscillators will
be done in section 3. Then, in section 4, and using as a guideline the
results obtained for the family of oscillators before, we will discuss the
tomographic picture of a real scalar Klein--Gordon on a finite cavity. For
that we will consider the field as described by a countable ensemble of
harmonic oscillators and a family of states similar to coherent states for
harmonic oscillators will be analyzed. It will be shown that the tomographic
description of such states is equivalent to the original one. The
tomographic description of the field states for the continuum case will be
discussed in section 5 following similar lines and finally, the Liouville
field equation in tomographic form will be discussed in section 6.
Conclusions and further perspectives of this work are given in section 7.

\section{The tomographic picture of classical physical systems: an overview}

The states of a classical system with a finite numbers of degrees of freedom
are described by a probability density $\rho (\boldsymbol{\varpi} )$ on its
phase space $\boldsymbol{\varpi} \in \Omega $. The phase space carries a
canonical measure, the Liouville measure $\mu _{\mathrm{Liouville}}$ that in
canonical coordinates $(\boldsymbol{q},\boldsymbol{p}),\boldsymbol{q}%
=(q_{1},\ldots ,q_{n}),\boldsymbol{p}=(p_{1},\ldots p_{n})$, has the form $%
\mathrm{d}\mu _{\mathrm{Liouville}}(\boldsymbol{q},\boldsymbol{p})=\mathrm{d}%
^{n}q\mathrm{d}^{n}p=\mathrm{d}q_{1}\cdots \mathrm{d}q_{n}\mathrm{d}%
p_{1}\cdots \mathrm{d}p_{n}$. In the case that $\Omega $ is a domain in $%
\mathbb{R}^{2n}$, the classical center--of--mass tomogram $\mathcal{W}_{%
\mathrm{cm}}$ of the state $\rho $ is defined as the Radon transform of the
density $\rho $ and consists of the average of $\rho $ along affine
hyperplanes on phase space, i.e.,
\begin{equation}
\mathcal{W}_{\mathrm{cm}}(X,\boldsymbol{\mu },\boldsymbol{\nu }%
)=\int_{\Omega }\rho \left( \boldsymbol{q},\boldsymbol{p}\right) \delta (X-%
\boldsymbol{\mu }\cdot \boldsymbol{q}-\boldsymbol{\nu }\cdot \boldsymbol{p})%
\mathrm{d}^{n}q\mathrm{d}^{n}p,  \label{class_radon}
\end{equation}%
where $\boldsymbol{\mu }=(\mu _{1},\ldots ,\mu _{n})$, $\boldsymbol{\nu }%
=(\nu _{1},\ldots \nu _{n})$ and the equation $X-\boldsymbol{\mu }\cdot
\boldsymbol{q}-\boldsymbol{\nu }\cdot \boldsymbol{p}=0$ determines an
hyperplane $\Pi $ in $\Omega $. The classical center--of--mass tomogram $%
\mathcal{W}_{\mathrm{cm}}(X,\boldsymbol{\mu },\boldsymbol{\nu })$ defines a
probability density, depending on the random variable $X,$ on the space of
hyperplanes in $\Omega $. The state $\rho $ can be reconstructed by using
the inverse Radon transform:
\begin{equation}
\rho \left( \boldsymbol{q},\boldsymbol{p}\right) =\int_{\mathbb{R}^{2n+1}}%
\mathcal{W}_{\mathrm{cm}}(X,\boldsymbol{\mu },\boldsymbol{\nu })\exp \left[
\mathrm{i}(X-\boldsymbol{\mu }\cdot \boldsymbol{q}-\boldsymbol{\nu }\cdot
\boldsymbol{p})\right] \mathrm{d}X\frac{\mathrm{d}^{n}\mu \mathrm{d}^{n}\nu
}{(2\pi )^{2n}}.  \label{inver_radon}
\end{equation}
where $\mathrm{d}^{n}\mu \mathrm{d}^{n}\nu=\mathrm{d}\mu _{1}\ldots \mathrm{d}\mu _{n}\mathrm{d}\nu _{1}\ldots \mathrm{d}\nu _{n}.$

Previous ideas can be extended by considering with more care the role of
the observables of the system in the construction of the tomogram $\mathcal{W%
}_{\mathrm{cm}}$. The description of a physical system involves always the
selection of its algebra of observables, call it $\mathcal{O}$, and its
corresponding states, denoted by $\mathcal{S}$. The outputs of measuring a
given observable $A\in \mathcal{O}$ when the system is in the state $\rho $
are described by a probability measure $\mu _{A,\rho }$ on the real line
such that $\mu _{A,\rho }(\Delta )$ is the probability that the output of $A$
belongs to the subset $\Delta \subset \mathbb{R}$. Thus a measure theory for
the physical system under consideration is a pairing between observables $A$
and states $\rho $ assigning to pairs of them measures $\mu _{A,\rho }$. In
this setting the expected value of the observable $A$ in the state $\rho $
is given simply by an integral over the real line parametrized by $\lambda$:
\begin{equation}
\langle A\rangle _{\rho }=\int_{\mathbb{R}} \lambda \mathrm{d}\mu _{A,\rho }.
\end{equation}%
Such picture applies equally well to classical and quantum systems. Thus for
closed quantum systems the observables are described by self--adjoint
operators $A$ on a Hilbert space $\mathcal{H}$ while states are described as
density operators $\rho $ acting on such Hilbert space. The pairing above is
provided by the assignment of the measure $\mu _{A,\rho }=\mathrm{Tr}(\rho
E_{A})$ where $E_{A}$ denotes the projector--valued spectral measure
associated to the Hermitian operator $A$.

The description of a classical system whose phase space is $\Omega $ can be
easily established in these terms by considering that the algebra of
observables $\mathcal{O}$ is a class (large enough) of functions on $\Omega, $%
 and that the states of the system are normalized positive functionals on $%
\mathcal{O}$, thus for instance if $\mathcal{O}$ contains the algebra of
continuous functions on $\Omega $, states are probability measures on phase
space. If we assume that the phase space is originally equipped with a
measure $\mu $, for instance the Liouville measure $\mu _{\mathrm{Liouville}%
} $ in the case of mechanical systems, then we may restrict ourselves to the
statistical states considered by Boltzmann corresponding to probability
measures which are absolutely continuous with respect to the Liouville
measure, thus determined by probability densities $\rho (\boldsymbol{\varpi}
)$ on $\Omega $. We denote such space of states by $\mathcal{S}$ as before.
Given an observable $f(\boldsymbol{\varpi} )$ on $\Omega $, the pairing
between states and observables will be realized by assigning to the
observable $f$ its characteristic distribution $\rho _{f}(\lambda )$ with respect to
the probability measure $\rho (\boldsymbol{\varpi} )\mathrm{d}\mu (%
\boldsymbol{\varpi} )$, then the probability of finding the measured value
of the observable $f$ in the interval $\Delta $ is given by:
\begin{equation}
\int_{\Delta }\rho _{f}(\lambda )\mathrm{d}\lambda ,
\end{equation}%
and the expected value of $f$ on the state $\rho $ will be given by:
\begin{equation}
\langle f\rangle _{\rho }=\int_{\mathbb{R}} \lambda \rho _{f}(\lambda )\mathrm{d}\lambda .
\end{equation}

The tomographic description provided by the classical center--of--mass
tomograms $\mathcal{W}_{\mathrm{cm}}$ above (\ref{class_radon}) does not
allow to cope with systems whose phase space is not of the previous form
(for instance spin systems) and it is convenient to expand the scope of the
formalism to make it more flexible and allow for alternative and more
general pictures. Other tomographic pictures have been proposed for both
classical and quantum systems (see for instance \cite {asorey}, and \cite{Ib11} for a
description of quantum tomograms in the realm of $C^{\ast }$--algebras).

A general tomographic picture of a classical system may be given starting with
a family of elements in $\mathcal{O}$ parametrized by an index $x$ which can
be discrete or continuous. Often $x$ is a point on a finite dimensional
manifold that we will denote by $\mathcal{M}$, thus $x\in \mathcal{M}$. We
will denote the observable associated to the element $x$ by $U(x)$ or $U_{x}$
depending on the context. Given a state $\rho $ of the system, the
correspondence $x\mapsto U_{x}$, allows to pull--back the observables $U_{x}$
to $\mathcal{M}$ defining the function $F_{\rho }(x)$ on $\mathcal{M}$
associated to the state $\rho (\boldsymbol{\varpi} )$ by:
\begin{equation}
F_{\rho }(x)=\langle \rho ,U(x)\rangle :=\int_{\Omega }U_{x}(\boldsymbol{\varpi} )\rho (\boldsymbol{%
\varpi} )\mathrm{d}\mu (\boldsymbol{\varpi} ).
\end{equation}%
The observables $U_{x}$ must be properly chosen so that previous integral is
defined. For instance we could have chosen $\mathcal{M}=\Omega $ as in the
definition of the Radon transform above (\ref{class_radon}), and then
consider $U_{\boldsymbol{\varpi} }=\delta (\boldsymbol{\varpi} )$, thus the
function $F_{\rho }$ associated to the state $\rho (\boldsymbol{\varpi} )$
will be again $\rho (\boldsymbol{\varpi} )$ itself. The original state $\rho
(\boldsymbol{\varpi} )$ could be reconstructed from $F_{\rho }$ iff the
family of observables $U(x)$ separate states, that is, given $\rho \neq \rho
^{\prime }$ two different states, there exists $x\in \mathcal{M}$ such that $%
F_{\rho }(x)=\langle \rho ,U(x)\rangle \neq \langle \rho ^{\prime
},U(x)\rangle =F_{\rho ^{\prime }}(x)$. Then two states are different if and
only if the corresponding representing functions $F_{\rho }$ are different.

Clearly up to now, our construction does not discriminate the description of
classical systems from quantum systems. The difference will appear only at
the level of the product structure on the induced functions $F_\rho,$ as the Wigner--Weyl--Moyal approach shows.
Another important ingredient for the tomographic description is the Radon
transform. To give an abstract presentation of this transform, we shall
assume for the time being that $\mathcal{M}$ is a manifold which carries a
measure, so that we can consider integrable functions on it and perform the
corresponding integrals.

Consider now the dual space of $\mathcal{F}(\mathcal{M}),$ denoted as $\mathcal{F}(\mathcal{M})^{\prime },$ and a second
auxiliary space $\mathcal{N}$ whose points will be denoted by $y\in \mathcal{%
N}$. The space $\mathcal{N}$ parametrizes a certain subspace $\mathcal{D}(%
\mathcal{M})\subset \mathcal{F}(\mathcal{M})^{\prime }$. In other words, for
each $y\in \mathcal{N}$ there is an assignment $y\mapsto D(y)$ with $D(y)\in
\mathcal{D}(\mathcal{M})$ a linear functional on the space of functions on $%
\mathcal{M}$. We obtain a map from $\mathcal{F}(\mathcal{M})$ to $\mathcal{F}%
(\mathcal{N})$ by setting for each $f\in \mathcal{F}(\mathcal{M})$:
\begin{equation}
\mathcal{W}_{f}(y)=\langle D(y),f\rangle .
\end{equation}

For instance suppose that $\mathcal{N}$ parametrizes a family of
submanifolds $S(y)$ of $\Omega $, $y\in \mathcal{N}$. If the submanifold $%
S(y)$ has the form $\Phi (\boldsymbol{q},\boldsymbol{p};X_{1},\ldots
,X_{d})=X_{0}$, $y=(X_{0},X_{1},\ldots ,X_{d})$ denoting a parametrization
of $\mathcal{N}$, the corresponding generalized Radon transform would be
written as:
\begin{equation}
\mathcal{W}(y)=\int_{\Omega }\rho (\boldsymbol{q},\boldsymbol{p})\delta
(X_{0}-\Phi (\boldsymbol{q},\boldsymbol{p};X_{1},\ldots ,X_{n}))\mathrm{d}%
^{n}q\mathrm{d}^{n}p,  \label{gen_radon}
\end{equation}%
which has the same form as eq. (\ref{class_radon}).

When the imbedding is properly chosen, it turns out that $\mathcal{W}(y)$ is
a fair probability distribution on $\mathcal{N}$ which we have constructed
out of the initial state $\rho $. The aim of tomography is to reconstruct $%
\rho $ out of the experimental distribution functions that we obtain from
the measurement of the selected observables parametrized by $\mathcal{M}$.
This is the so called inversion formula for the Radon transform. In the case
that $\Omega =\mathbb{R}^{2n}$ and $\mathcal{N}$ denotes as in (\ref%
{class_radon}) the space of hyperplanes, then because of the homogeneity
properties of the Dirac distribution, we find that $\mathcal{W}_{\mathrm{cm}%
} $ satisfies the condition:
\begin{equation}
\left[ X\frac{\partial }{\partial X}+\boldsymbol{\mu }\cdot \frac{\partial }{%
\partial \boldsymbol{\mu }}+\boldsymbol{\nu }\cdot \frac{\partial }{\partial
\boldsymbol{\nu }}+1\right] \mathcal{W}_{\mathrm{cm}}(X,\boldsymbol{\mu },%
\boldsymbol{\nu })=0.  \label{homogeneity}
\end{equation}%
Due to the homogeneity condition (\ref{homogeneity}), $\mathcal{W}_{\mathrm{%
cm}}$ depends effectively only on $2n$ variables instead of $2n+1$ and the
inversion formula works, out of the \textquotedblleft
measurements\textquotedblright\ performed with the family of observables $\{%
\boldsymbol{\mu }\cdot \boldsymbol{\hat{q}}+\boldsymbol{\nu }\cdot
\boldsymbol{\hat{p}}\},(\boldsymbol{\mu },\boldsymbol{\nu })\in \mathbb{R}%
^{2n},$ we are able to recover $\rho $ by means of eq. (\ref{inver_radon}).

As an important example of the previous discussion, we introduce another
kind of tomographic representation of the state $\rho (\boldsymbol{q},%
\boldsymbol{p}),$ the classical symplectic tomogram defined as:
\begin{equation}
\mathcal{W}_{\rho }(\boldsymbol{X},\boldsymbol{\mu },\boldsymbol{\nu }%
)=\int_{\mathbb{R}^{2n}}\rho (\boldsymbol{q},\boldsymbol{p}%
)\prod_{k=1}^{n}\delta (X_{k}-\mu _{k}q_{k}-\nu _{k}p_{k})\mathrm{d}%
q_{1}\ldots \mathrm{d}q_{n}\mathrm{d}p_{1}\ldots \mathrm{d}p_{n}.
\end{equation}%
Notice that we have taken $\mathcal{M}=\mathbb{R}^{2n}$, the phase space
again, and $\mathcal{N}=\mathcal{N}_{1}\times \cdots \times \mathcal{N}_{n}$
with $\mathcal{N}_{k}$ the space of lines in $\mathbb{R}^{2}$, the phase
space of each individual degree of freedom of the physical system under
consideration. Thus, we have obtained a joint probability distribution of
the $n$ random variables $\left( X_{1},\dots ,X_{n}\right) =\boldsymbol{X}.$
In contrast to the center--of--mass case, because of the presence of $n$
Dirac distributions, we find that the symplectic tomogram $\mathcal{W}_{\rho
}$ satisfies $n$\ homogeneity conditions:
\begin{equation}
\left[ X_{k}\frac{\partial }{\partial X_{k}}+\mu _{k}\frac{\partial }{%
\partial \mu _{k}}+\nu _{k}\frac{\partial }{\partial \nu _{k}}+1\right]
\mathcal{W}_{\rho }(\boldsymbol{X},\boldsymbol{\mu },\boldsymbol{\nu }%
)=0\quad ,\quad k=1,\dots ,n.
\end{equation}%
In other words, the classical symplectic tomogram $\mathcal{W}_{\rho }(%
\boldsymbol{X},\boldsymbol{\mu },\boldsymbol{\nu })$\ depends effectively
only on $2n$ variables instead of $3n.$ In fact, one can show that the
symplectic tomogram $\mathcal{W}_{\rho }(\boldsymbol{X},\boldsymbol{\mu },%
\boldsymbol{\nu })$ can be transformed into the center--of--mass tomogram $%
\mathcal{W}_{\mathrm{cm}}$ of the same state $\rho ,$ and \textit{vice versa}%
. Finally, out of the \textquotedblleft measurements\textquotedblright\
performed with the family of observables $\{\mu _{k}\hat{q}_{k}+\nu _{k}\hat{p}%
_{k}\},(\mu _{k},\nu _{k})\in \mathbb{R}^{2},k=1,\dots ,n,$ we are again
able to recover $\rho $, by means of the symplectic inversion formula:%
\begin{equation}
\rho (\boldsymbol{q},\boldsymbol{p})=\int_{\mathbb{R}^{3n}}\mathcal{W}_{\rho
}(\boldsymbol{X},\boldsymbol{\mu },\boldsymbol{\nu })\exp \left[ \mathrm{i}%
\sum_{k=1}^{n}(X_{k}-\mu _{k}q_{k}-\nu _{k}p_{k})\right] \mathrm{d}^{n}X%
\frac{\mathrm{d}^{n}\mu \mathrm{d}^{n}\nu }{(2\pi )^{2n}},
\label{inversympl}
\end{equation}%
where $\mathrm{d}^{n}X=\mathrm{d}X_{1}\dots \mathrm{d}X_{n}.$

\section{Tomograms for states of an ensemble of classical oscillators}

\subsection{The canonical ensemble}

If we consider a family of $n$ independent one--dimensional oscillators with
frequencies $\omega _{k}>0$, its phase space $\Omega $ will be $\mathbb{R}%
^{2n}$ with canonical coordinates $(q_{k},p_{k})$, $k=1,\ldots ,n$. The
Hamiltonian of the system will be $H=\sum_{k=1}^{n}H_{k}$, $%
H_{k}(q_{k},p_{k})=\frac{1}{2}(p_{k}^{2}+\omega _{k}^{2}q_{k}^{2})$. The
dynamics of the system will be given by
\begin{equation}
\dot{q}_{k}=p_{k},\quad \dot{p}_{k}=-\omega _{k}^{2}q_{k},\quad k=1,\ldots n,
\label{standard}
\end{equation}%
and the Liouville measure on phase space takes again the form $\mathrm{d}\mu
_{\mathrm{Liouville}}=\mathrm{d}q_{1}\ldots \mathrm{d}q_{n}\mathrm{d}%
p_{1}\ldots \mathrm{d}p_{n}$. Making the change of variables $\xi _{k}=q_{k}/%
\sqrt{\omega _{k}}$, $\eta _{k}=\sqrt{\omega _{k}}p_{k}$, the dynamics is
written in the symmetrical form
\begin{equation}
\dot{\xi}_{k}=\omega _{k}\eta _{k},\quad \dot{\eta}_{k}=-\omega _{k}\xi
_{k}.\quad k=1,\ldots n.  \label{symm_dyn}
\end{equation}%
and the Hamiltonian becomes
\begin{equation}
H(\xi ,\eta )=\sum_{k=1}^{n}H_{k}(\xi_{k} ,\eta_{k} )=\frac{1}{2}%
\sum_{k=1}^{n}\omega _{k}(\xi _{k}^{2}+\eta _{k}^{2}).  \label{ham_osci}
\end{equation}%
The Liouville measure remains unchanged under this change of variables $%
\mathrm{d}\mu _{\mathrm{Liouville}}(\boldsymbol{q},\boldsymbol{p})=\mathrm{d}%
^{n}q\mathrm{d}^{n}p=\mathrm{d}^{n}\xi \mathrm{d}^{n}\eta =\mathrm{d}\xi
_{1}\ldots \mathrm{d}\xi _{n}\mathrm{d}\eta _{1}\ldots \mathrm{d}\eta _{n}=%
\mathrm{d}\mu _{\mathrm{Liouville}}(\boldsymbol{\xi },\boldsymbol{\eta })$
and statistical states are described by probability densities $\rho (%
\boldsymbol{q},\boldsymbol{p})=\rho (\boldsymbol{\xi },\boldsymbol{\eta })$.
Liouville equation determines the evolution of the state:
\begin{equation}
\frac{\mathrm{d} }{\mathrm{d} t}\rho =\{\rho ,H\}  \label{Liouville_class}
\end{equation}%
where the Poisson brackets are defined by the canonical commutation
relations $\{q_{k},p_{l}\}=\delta _{kl}$, $\{q_{k},q_{l}\}=\{p_{k},p_{l}\}=0$%
. Notice that if $\boldsymbol{\varpi} =(\boldsymbol{\xi },\boldsymbol{\eta }%
)\in \Omega $ is a point in phase space, then $\rho _{t}(\boldsymbol{\varpi}
)=\rho (\boldsymbol{\xi }(t),\boldsymbol{\eta }(t))$ with $\left(
\boldsymbol{\xi }(t),\boldsymbol{\eta }(t)\right) $ the solution of the
equations of motion (\ref{symm_dyn}) starting at $\boldsymbol{\varpi} $ at
time $t=0$.

In particular, the Gibbs state or canonical distribution is given by $\rho _{%
\mathrm{can}}(\boldsymbol{q},\boldsymbol{p})=\mathrm{e}^{-\beta H}/Z_{0}$
where the normalization constant $Z_{0}$ is easily evaluated
\begin{eqnarray}
Z_{0} &=&\int_{\Omega }\mathrm{e}^{-\beta H(\boldsymbol{q},\boldsymbol{p})}%
\mathrm{d}\mu _{\mathrm{Liouville}}(\boldsymbol{q},\boldsymbol{p})=\int_{%
\mathbb{R}^{2n}}\mathrm{e}^{-\frac{1}{2}\beta \sum_{k=1}^{n}\omega _{k}(\xi
_{k}^{2}+\eta _{k}^{2})}\mathrm{d}\xi _{1}\ldots \mathrm{d}\xi _{n}\mathrm{d}%
\eta _{1}\ldots \mathrm{d}\eta _{n} \\
&=&\left( 2\pi \right) ^{n}\prod_{k=1}^{n}(\beta \omega _{k})^{-1}.  \notag
\end{eqnarray}

Hence for a given observable $f$ we will have:
\begin{equation}
\langle f\rangle _{\rho _{\mathrm{can}}}=\frac{1}{Z_{0}}\int_{\mathbb{R}%
^{2n}}f(\boldsymbol{\xi },\boldsymbol{\eta })\mathrm{e}^{-\frac{1}{2}\beta
\sum_{k=1}^{n}\omega _{k}(\xi _{k}^{2}+\eta _{k}^{2})}\mathrm{d}\xi
_{1}\ldots \mathrm{d}\xi _{n}\mathrm{d}\eta _{1}\ldots \mathrm{d}\eta _{n}.
\end{equation}
More detailed information will be found in \cite{sudklaudbook}.

The classical tomographic description of a state $\rho (\boldsymbol{\xi },%
\boldsymbol{\eta })$ will be performed by means of a symplectic tomogram:
\begin{equation}
\mathcal{W}_{\rho }(\mathbf{X},\boldsymbol{\mu },\boldsymbol{\nu })=\int_{%
\mathbb{R}^{2n}}\rho (\boldsymbol{\xi },\boldsymbol{\eta }%
)\prod_{k=1}^{n}\delta (X_{k}-\mu _{k}\xi _{k}-\nu _{k}\eta _{k})\mathrm{d}%
\xi _{1}\ldots \mathrm{d}\xi _{n}\mathrm{d}\eta _{1}\ldots \mathrm{d}\eta
_{n}.  \label{class_tom_oscillators}
\end{equation}%
We recall that here we have taken $\mathcal{M}=\mathbb{R}^{2n}$, the phase
space again, and $\mathcal{N}=\mathcal{N}_{1}\times \cdots \times \mathcal{N}%
_{n}$ with $\mathcal{N}_{k}$ the space of lines in $\mathbb{R}^{2}$, the
phase space of each individual one--dimensional oscillator. A simple
computation shows that the Gibbs state tomogram reads:
\begin{equation}
\mathcal{W}_{\rho _{\mathrm{can}}}(\boldsymbol{X},\boldsymbol{\mu },%
\boldsymbol{\nu })=\prod_{k=1}^{n}\sqrt{\frac{\beta \omega _{k}}{2\pi \left(
\mu _{k}^{2}+\nu _{k}^{2}\right) }}\exp \left[ -\frac{\beta \omega
_{k}X_{k}^{2}}{2(\mu _{k}^{2}+\nu _{k}^{2})}\right] .
\end{equation}

A interesting family of states which are the classical counterpart of
quantum coherent states can be introduced by means of the holomorphic
representation $\zeta _{k}=\frac{1}{\sqrt{2}}(\xi _{k}+i\eta _{k})$ of phase
space, hence the phase space becomes the complex space $\mathbb{C}^{n}$ with
the Hermitian structure $H(\boldsymbol{\zeta },\boldsymbol{\bar{\zeta}}%
)=\sum_{k=1}^{n}\omega _{k}|\zeta _{k}|^{2}$. Given a point $\boldsymbol{z}%
=(z_{1},\ldots ,z_{n})\in \mathbb{C}^{n}$ we can construct the distribution
\begin{equation}
\rho _{\boldsymbol{z}}(\boldsymbol{\zeta },\boldsymbol{\bar{\zeta}})=N\left(
\boldsymbol{z}\right) \exp \left[ \sum_{k=1}^{n}\omega _{k}(z_{k}\bar{\zeta}%
_{k}+\bar{z}_{k}\zeta _{k})\right] \left. \rho _{\mathrm{can}}(\boldsymbol{%
\zeta },\boldsymbol{\bar{\zeta}})\right\vert _{\beta =1}
\end{equation}%
where%
\begin{equation}
N\left( \boldsymbol{z}\right) =\prod_{k=1}^{n}\pi \omega _{k}^{-1}\mathrm{%
\exp }\left[ -\omega _{k}|z_{k}|^{2}\right] .
\end{equation}%
Notice that integrating the Liouville equation for such state yields:
\begin{equation}
\rho _{\boldsymbol{z}}\left( t\right) =\rho _{\boldsymbol{z}(t)},
\end{equation}%
with
\begin{equation}
z_{k}(t)=e^{-i\omega _{k}t}z_{k}(0).
\end{equation}%
The symplectic tomographic distribution corresponding to $\rho _{\boldsymbol{%
z}}(\boldsymbol{\zeta },\boldsymbol{\bar{\zeta}})$ is a product%
\begin{equation}
\mathcal{W}_{\rho _{\boldsymbol{z}}}(\boldsymbol{X},\boldsymbol{\mu },%
\boldsymbol{\nu },\boldsymbol{z})=\prod_{k=1}^{n}\mathcal{W}_{\rho _{%
\boldsymbol{z}}}^{\left( k\right) }(X_{k},\mu _{k},\nu _{k},z_{k}),
\end{equation}%
where the tomogram $\mathcal{W}_{\rho _{\boldsymbol{z}}}^{\left( k\right) }$%
\ of a single degree of freedom is a Gaussian distribution
\begin{equation}
\mathcal{W}_{\rho _{\boldsymbol{z}}}^{\left( k\right) }(X_{k},\mu _{k},\nu
_{k},z_{k})=\sqrt{\frac{\omega _{k}}{2\pi \left( \mu _{k}^{2}+\nu
_{k}^{2}\right) }}\exp \left[ -\frac{\omega _{k}\left( X_{k}-\left\langle
X_{k}\left( \mu _{k},\nu _{k},z_{k}\right) \right\rangle \right) ^{2}}{2(\mu
_{k}^{2}+\nu _{k}^{2})}\right]
\end{equation}%
of the random variable $X_{k},$ with mean value
\begin{equation}
\left\langle X_{k}\left( \mu _{k},\nu _{k},z_{k}\right) \right\rangle =\mu
_{k}\Re \left( z_{k}\right) +\nu _{k}\Im \left( z_{k}\right)
\end{equation}%
and variance given by%
\begin{equation}
\sigma _{X_{k}X_{k}}=\frac{2(\mu _{k}^{2}+\nu _{k}^{2})}{\omega _{k}}.
\end{equation}

\subsection{A new class of states: Gauss--Laguerre states}

We will introduce now a family of classical states, called Gauss--Laguerre (GL) states, inspired on the
Wigner functions of the excited states of a quantum harmonic oscillator. These functions are only quasi--distributions on phase space,
however their squares are related to the purity of the corresponding quantum states and are true probability distributions \cite{dodman}. Thus, the family of classical states we consider is defined as:
\begin{equation}
\rho_{\mathrm{GL},\left\{ m\right\} }(\boldsymbol{\xi },\boldsymbol{%
\eta })=\prod_{k=1}^{n}\rho_{%
\mathrm{GL},m_{k}}^{\left( k\right) }(\xi _{k},\eta _{k}),
\end{equation}%
where $\left\{ m\right\} =\left\{
m_{1},m_{2},\dots ,m_{n}\right\} $ is a multi--index and
\begin{equation}
\rho_{\mathrm{GL},m_{k}}^{\left( k\right) }(\xi _{k},\eta _{k})=\frac{%
\omega _{k}}{2\pi }\left[ L_{m_{k}}\left( \frac{\omega _{k}}{2}(\xi
_{k}^{2}+\eta _{k}^{2})\right) \right] ^{2}\mathrm{e}^{-\frac{1}{2}\omega
_{k}(\xi _{k}^{2}+\eta _{k}^{2})}.
\end{equation}%
Here $L_{m_{k}}$ is the Laguerre
polynomial of degree $m_{k}$ and the Gaussian exponential is the not normalized Gibbs
state $\varrho _{\mathrm{can}}|_{\beta =1}$. Notice that $\rho_{%
\mathrm{GL},m_{k}}^{\left( k\right) }(\xi _{k},\eta _{k})$ is a classical state
on a bidimensional phase space.

The symplectic Radon transform of the state factorizes:
\begin{equation}
\mathcal{W}_{\mathrm{GL},\left\{ m\right\} }(\boldsymbol{X},%
\boldsymbol{\mu },\boldsymbol{\nu })=\prod_{k=1}^{n}\mathcal{W}_{\mathrm{GL}%
,m_{k}}^{\left( k\right) }(X_{k},\mu _{k},\nu _{k}).
\end{equation}%
and we will obtain%
\begin{equation}
\mathcal{W}_{\mathrm{GL},m_{k}}^{\left( k\right) }(X_{k},\mu _{k},\nu _{k})=%
\frac{\mathrm{\exp }\left[ -\frac{X_{k}^{2}}{\sigma _{k}^{2}}\right] }{\pi ^{%
\frac{1}{2}}\sigma _{k}}\sum_{s=0}^{m_{k}}\frac{1}{2^{2m_{k}}}\binom{2\left(
m_{k}-s\right) }{m_{k}-s}\binom{2s}{s}\frac{\left[ H_{2s}\left( \frac{X_{k}}{%
\sigma _{k}}\right) \right] }{2^{2s}\left( 2s\right) !}^{2}
\end{equation}%
with%
\begin{equation}
\sigma _{k}=\sqrt{\frac{2\left( \mu _{k}^{2}+\nu _{k}^{2}\right) }{\omega
_{k}}},
\end{equation}%
while $H_{2s}$ is the Hermite polynomial of degree $2s.$ The above result
can be obtained as follows.

First, we drop the label $k$ and write $\mathcal{W}_{m}(X,\mu ,\nu )$ in
place of $\mathcal{W}_{\mathrm{GL},m_{k}}^{\left( k\right) }(X_{k},\mu
_{k},\nu _{k}).$ Thus%
\begin{eqnarray}
\mathcal{W}_{m}(X,\mu ,\nu ) &=&\frac{\omega }{2\pi }\int L_{m}^{2}\left(
\frac{\omega }{2}(\xi ^{2}+\eta ^{2})\right) \mathrm{e}^{-\frac{1}{2}\omega
(\xi ^{2}+\eta ^{2})} \delta \left( X-\mu \xi -\nu \eta \right) \mathrm{d}%
\xi \mathrm{d}\eta \\
&=&\frac{\omega }{\left( 2\pi \right) ^{2}}\int \mathrm{d}K\mathrm{e}^{%
\mathrm{i}KX}\int L_{m}^{2}\left( \frac{\omega }{2}(\xi ^{2}+\eta
^{2})\right) \mathrm{e}^{-\frac{1}{2}\omega (\xi ^{2}+\eta ^{2})} \mathrm{e}%
^{-\mathrm{i}K\left( \mu \xi +\nu \eta \right) }\mathrm{d}\xi \mathrm{d}\eta
.  \notag
\end{eqnarray}

Now we put $\sqrt{\mu ^{2}+\nu ^{2}}=r_{\mu \nu },\mu =r_{\mu \nu }\cos
\alpha _{\mu \nu },\nu =r_{\mu \nu }\sin \alpha _{\mu \nu },$ and $\xi
=r\sin \theta ,\eta =r\cos \theta .$ Then, we recast the previous formula as%
\begin{equation}
\mathcal{W}_{m}(X,\mu ,\nu )=\frac{1}{2\pi }\int \mathrm{d}K\mathrm{e}^{%
\mathrm{i}KX}\mathcal{\tilde{W}}_{m}(K,\mu ,\nu )
\end{equation}%
where the characteristic function of $\mathcal{W}_{m},$ i.e. its Fourier
transform $\mathcal{\tilde{W}}_{m},$ is given by
\begin{equation}
\mathcal{\tilde{W}}_{m}(K,\mu ,\nu )=\int_{0}^{2\pi }\frac{\mathrm{d}\theta
}{2\pi }\int_{0}^{\infty }\left[ L_{m}\left( \frac{\omega r^{2}}{2}\right) %
\right] ^{2}\mathrm{e}^{-\frac{\omega r^{2}}{2}}\mathrm{e}^{^{-\mathrm{i}%
\left( Kr_{\mu \nu }\right) r\sin \left( \theta +\alpha _{\mu \nu }\right) }}%
\mathrm{d}\left( \frac{\omega r^{2}}{2}\right) .
\end{equation}%
The integral over the angular variable $\theta _{\mu \nu }=\theta +\alpha
_{\mu \nu }$ yields the Bessel function $J_{0}$, so:%
\begin{equation}
\mathcal{\tilde{W}}_{m}(K,\mu ,\nu )=\int_{0}^{\infty }\left[ L_{m}\left(
\frac{x^{2}}{2}\right) \right] ^{2}\mathrm{e}^{-\frac{x^{2}}{2}}J_{0}\left(
\frac{Kr_{\mu \nu }}{\sqrt{\omega }}x\right) \mathrm{d}\left( \frac{x^{2}}{2}%
\right) .
\end{equation}%
The above integral can be evaluated and gives (\cite{GraRyz}, n. 7.422 2)%
\begin{eqnarray}
\mathcal{\tilde{W}}_{m}(K,\mu ,\nu ) &=&\mathrm{e}^{-\frac{1}{2}\left( \frac{%
Kr_{\mu \nu }}{\sqrt{\omega }}\right) ^{2}}\left[ L_{m}\left( \frac{1}{2}%
\left( \frac{Kr_{\mu \nu }}{\sqrt{\omega }}\right) ^{2}\right) \right] ^{2}
\label{LaguerreChar} \\
&=&\mathrm{e}^{-\frac{1}{2}\left( \frac{Kr_{\mu \nu }}{\sqrt{\omega }}%
\right) ^{2}}\frac{1}{2^{2m}}\sum_{s=0}^{m}\binom{2\left( m-s\right) }{m-s}%
\binom{2s}{s}L_{2s}\left( \left( \frac{Kr_{\mu \nu }}{\sqrt{\omega }}\right)
^{2}\right),  \notag
\end{eqnarray}%
where the last line has been obtained by a well known addition formula of
Laguerre polynomials (\cite{GraRyz}, n. 8.976 3).

We remark that the above equation yields, by multiplication over the
restored label $k,$ the characteristic function $\mathcal{\tilde{W}}_{%
\mathrm{GL},\left\{ m\right\} }(\boldsymbol{K},\boldsymbol{\mu },%
\boldsymbol{\nu })$ , with $\boldsymbol{K=}\left( K_{1},\dots ,K_{k},\dots
K_{n}\right) $, of the tomogram $\mathcal{W}_{\mathrm{GL},\left\{
m\right\} }(\boldsymbol{X},\boldsymbol{\mu },\boldsymbol{\nu }).$

Besides, as $\mathcal{\tilde{W}}_{m}(K=0,\mu ,\nu )=1,$ we get at once the
normalization property of the tomogram $\mathcal{W}_{m}(X,\mu ,\nu ).$

Finally, we are able to perform the last integration. The Fourier
anti--transform of $\mathcal{\tilde{W}}_{m}(K,\mu ,\nu )$\ is obtained by
means of the integral over $y=Kr_{\mu \nu }/\sqrt{\omega }$ (\cite{GraRyz},
n. 7.418 2):
\begin{eqnarray}
\frac{1}{\pi }\frac{\sqrt{\omega }}{r_{\mu \nu }}\int_{0}^{\infty
}L_{2s}\left( y^{2}\right) \mathrm{e}^{-\frac{1}{2}y^{2}}\cos \left( \frac{%
\sqrt{\omega }}{r_{\mu \nu }}Xy\right) \mathrm{d}y =\frac{\sqrt{\omega }}{%
\sqrt{2\pi }r_{\mu \nu }}\mathrm{e}^{-\frac{\omega }{2\left( r_{\mu \nu
}\right) ^{2}}X^{2}}\frac{1}{2^{2s}\left( 2s\right) !}\left[ H_{2s}\left(
\frac{\sqrt{\omega }}{\sqrt{2}r_{\mu \nu }}X\right) \right] ^{2}.
\end{eqnarray}%
So, we get the predicted expression of $\mathcal{W}_{m}(X,\mu ,\nu ).$

\section{The tomographic picture of Liouville's equation}

\label{harmonic} Finally, let us discuss the tomographic form of the
evolution equation for states, Liouville equation (\ref{Liouville_class}).
The evolution equation in the tomographic description was recently obtained
in \cite{Ch07} in relation with a relativistic wave function description of
harmonic oscillators. We will describe it here in the realm of our previous
discussion. Notice that because of the symplectic reconstruction formula for
a classical state (\ref{inversympl}) we can compute:
\begin{equation}
\frac{\partial }{\partial t}\rho (\boldsymbol{\xi },\boldsymbol{\eta }%
,t)=\int_{\mathbb{R}^{3n}}\exp \left[ \mathrm{i}\sum_{k=1}^{n}(X_{k}-\mu
_{k}\xi _{k}-\nu _{k}\eta _{k})\right] \frac{\partial }{\partial t}\mathcal{W%
}_{\rho }(\boldsymbol{X},\boldsymbol{\mu },\boldsymbol{\nu },t)\mathrm{d}%
^{n}X\frac{\mathrm{d}^{n}\mu \mathrm{d}^{n}\nu }{(2\pi )^{2n}},
\end{equation}%
(notice that the symplectic tomogram is computed at a given fixed time) and,
on the other hand:
\begin{eqnarray}
&&\{\rho ,H\} =\sum_{k=1}^{n}\left[ \frac{\partial H}{\partial \eta _{k}}%
\frac{\partial }{\partial \xi _{k}}-\frac{\partial H}{\partial \xi _{k}}%
\frac{\partial }{\partial \eta _{k}}\right] \rho \\
&=&\sum_{k=1}^{n}\int_{\mathbb{R}^{3n}}\mathrm{d}^{n}X\frac{\mathrm{d}%
^{n}\mu \mathrm{d}^{n}\nu }{(2\pi )^{2n}}\mathcal{W}_{\rho }(\boldsymbol{X},%
\boldsymbol{\mu },\boldsymbol{\nu },t) \left[ \frac{\partial H}{\partial
\eta _{k}}\frac{\partial }{\partial \xi _{k}}-\frac{\partial H}{\partial \xi
_{k}}\frac{\partial }{\partial \eta _{k}}\right] \exp \left[ \mathrm{i}%
\sum_{j=1}^{n}(X_{j}-\mu _{j}\xi _{j}-\nu _{j}\eta _{j})\right]  \notag \\
&=&\sum_{k=1}^{n}\int_{\mathbb{R}^{3n}}\mathrm{d}^{n}X\frac{\mathrm{d}%
^{n}\mu \mathrm{d}^{n}\nu }{(2\pi )^{2n}}\mathcal{W}_{\rho }(\boldsymbol{X},%
\boldsymbol{\mu },\boldsymbol{\nu },t) \,\left[ \frac{\partial H}{\partial
\xi _{k}}\nu _{k}\frac{\partial }{\partial X_{k}}-\frac{\partial H}{\partial
\eta _{k}}\mu _{k}\frac{\partial }{\partial X_{k}}\right] \exp \left[
\mathrm{i}\sum_{j=1}^{n}(X_{j}-\mu _{j}\xi _{j}-\nu _{j}\eta _{j})\right] .
\notag
\end{eqnarray}%
Eventually, we obtain the evolution equation for the classical tomogram $%
\mathcal{W_{\rho }}$:
\begin{eqnarray}
&&\frac{\partial \mathcal{W}_{\rho }(\boldsymbol{X},\boldsymbol{\mu },%
\boldsymbol{\nu },t)}{\partial t}= \\
&&\sum_{k=1}^{n}\left[ \frac{\partial H}{\partial \eta _{k}}\left( \left\{
\xi _{j}\rightarrow -\left[ \frac{\partial }{\partial X_{j}}\right] ^{-1}%
\frac{\partial }{\partial \mu _{j}}\right\} ,\left\{ \eta _{j}\rightarrow %
\left[ \frac{\partial }{\partial X_{j}}\right] ^{-1}\frac{\partial }{%
\partial \nu _{j}}\right\} \right) \mu _{k}\frac{\partial }{\partial X_{k}}%
\mathcal{W}_{\rho }(\boldsymbol{X},\boldsymbol{\mu },\boldsymbol{\nu }%
,t)\right.  \notag \\
&&\left. -\frac{\partial H}{\partial \xi _{k}}\left( \left\{ \xi
_{j}\rightarrow -\left[ \frac{\partial }{\partial X_{j}}\right] ^{-1}\frac{%
\partial }{\partial \mu _{j}}\right\} ,\left\{ \eta _{j}\rightarrow \left[
\frac{\partial }{\partial X_{j}}\right] ^{-1}\frac{\partial }{\partial \nu
_{j}}\right\} \right) \nu _{k}\frac{\partial }{\partial X_{k}}\mathcal{W}%
_{\rho }(\boldsymbol{X},\boldsymbol{\mu },\boldsymbol{\nu },t)\right] .
\notag
\end{eqnarray}%
Notice that the arguments $\left\{ \xi _{j}\right\} ,\left\{ \eta
_{j}\right\} $ of the derivatives of \ $H,$\ for any $j,$ are replaced by
the operators $\left\{ -\left[ \frac{\partial }{\partial X_{j}}\right] ^{-1}%
\frac{\partial }{\partial \mu _{j}}\right\} ,\left\{ \left[ \frac{\partial }{%
\partial X_{j}}\right] ^{-1}\frac{\partial }{\partial \nu _{j}}\right\} ,$
respectively. Explicitly, the operator $\left[ \frac{\partial }{\partial X}%
\right] ^{-1}$ is defined in terms of a Fourier transform as%
\begin{equation}
\left[ \frac{\partial }{\partial X}\right] ^{-1}\int_{\mathbb{R}}f\left(
K\right) \exp \left( \mathrm{i}KX\right) dK=\int_{\mathbb{R}}\frac{f\left(
K\right) }{\mathrm{i}K}\exp \left( \mathrm{i}KX\right) \mathrm{d}K.
\end{equation}%
Due to the presence of such terms, for a generic Hamiltonian $H$\ the
evolution tomographic equation is integro-differential. In the particular
instance of $H$ given by (\ref{ham_osci}), because of the general
correspondence rule:%
\begin{equation}
\frac{\partial }{\partial \xi _{k}}\rho \leftrightarrow \mu _{k}\frac{%
\partial }{\partial X_{k}}\mathcal{W}_{\rho }\quad ,\quad \frac{\partial }{%
\partial \eta _{k}}\rho \leftrightarrow \mu _{k}\frac{\partial }{\partial
X_{k}}\mathcal{W}_{\rho },
\end{equation}%
the tomographic evolution equation takes the form of a differential
equation:
\begin{eqnarray}  \label{evol_tomo_class}
&&\frac{\partial \mathcal{W}_{\rho }(\boldsymbol{X},\boldsymbol{\mu },%
\boldsymbol{\nu },t)}{\partial t}=\sum_{k=1}^{n}\omega _{k}\left[ \mu _{k}%
\frac{\partial }{\partial \nu _{k}}-\nu _{k}\frac{\partial }{\partial \mu
_{k}}\right] \mathcal{W}_{\rho }(\boldsymbol{X},\boldsymbol{\mu },%
\boldsymbol{\nu },t)  \label{evol_tomogram_osci} \\
&=&\sigma \left( \left\{ \mu _{k},\nu _{k}\right\} _{k},\left\{ \xi
_{k}\rightarrow \omega _{k}\frac{\partial }{\partial \mu _{k}},\eta
_{k}\rightarrow \omega _{k}\frac{\partial }{\partial \nu _{k}}\right\}
_{k}\right) \mathcal{W}_{\rho }(\boldsymbol{X},\boldsymbol{\mu },\boldsymbol{%
\nu },t),  \notag
\end{eqnarray}%
where $\sigma $ is the canonical symplectic form on the linear space $E=%
\mathbb{R}^{2n}$.

\section{The tomogram of the real Klein-Gordon field in a cavity}

Having shown that an interesting family of states for a finite ensemble of
harmonic oscillators is amenable to be described tomographically, we will
discuss now the Klein--Gordon equation for a real scalar field $\varphi (%
\mathrm{x})$ in a finite cavity on $1+d$ Minkowski space--time. Thus we
consider Minkowski space--time $\mathbb{M}=\mathbb{R}^{1+d}$ with metric of
signature $(+,-,\cdots ,-)$. Points in space--time will be written as $%
\mathrm{x}=(t,x)$ The dynamics of the real scalar field $\varphi (\mathrm{x}%
)=\varphi (t,x)$ is defined by the Lagrangian density:
\begin{equation}
\mathcal{L}\left[ \varphi \right] =\frac{1}{2}\left( \partial _{\mu }\varphi
\partial ^{\mu }\varphi -V\left[ \varphi \right] \right) ,
\end{equation}%
with Euler--Lagrange equations:
\begin{equation}
\partial _{\mu }\partial ^{\mu }\varphi =-V^{\prime }\left[ \varphi \right] .
\end{equation}%
Considering $V\left[ \varphi \right] =m^{2}\varphi ^{2}$ we get the
Klein--Gordon equation:
\begin{equation}
\varphi _{tt}-\Delta \varphi +m^{2}\varphi =0,
\end{equation}%
with $\Delta $ the $d$--dimensional Laplacian in $\mathbb{R}^{d}$. As we have extensively seen,
tomographic methods are described on phase space where conjugated variables and Poisson brackets are available.
On this carrier space dynamical equations are described by a vector field, first order differential equations in time.
Thus, for our Klein--Gordon equations we have to introduce a larger carrier space where the equations will be first order
in time. The transition from second order equations in time to first order differential equations in time may be done in many ways \cite{masasivibook}, here we shall consider one in which the new variables will make the equations of motion more symmetric.
We would stress that by using a specific splitting of spacetime into a space part and a time part we break the explicit Poincar\`{e}
invariant form but of course our description is still relativistic invariant.
 To proceed, we will
consider the Cauchy hypersurface $\mathcal{C}=\{0\}\times \mathbb{R}^{d}$
and the finite cavity will be defined as $\mathcal{V}\subset \mathcal{C}$.
We consider the restriction of the field to the cavity $\mathcal{V}$ using
the same notation $\varphi (x):=\varphi (0,x),x\in \mathcal{V}$ and the
Klein--Gordon equation becomes the evolution equation in the space of fields
$\varphi (x)$:
\begin{equation}
\frac{\mathrm{d}^2 \varphi}{\mathrm{d} t^2}=-(-\Delta +m^{2})\varphi .  \label{KG}
\end{equation}%
Boundary conditions at the boundary of the cavity $\mathcal{V}$ are chosen
such that the operator $-\Delta +m^{2}$ is strictly--positive and
self-adjoint on square integrable functions on $\mathcal{V}$ with respect to
the Lebesgue measure, thus we can define the invertible positive
self--adjoint operator $B=\sqrt{-\Delta +m^{2}}$. We will also assume for
simplicity that boundary conditions are chosen in such a way that the
spectrum of $B$ is nondegenerate, so that the eigenvalues of $B$ will be $%
0<\omega _{1}<\omega _{2}<\ldots <\omega _{n}<\ldots $ with eigenfunctions $%
\Phi _{k}(x)$, $B\Phi _{k}(x)=\omega _{k}\Phi _{k}(x)$, $k=1,2,\ldots $.
Thus equation (\ref{KG}) may be transformed into a first order evolution
differential equation by introducing the new fields:
\begin{equation}
\xi =B^{1/2}\varphi \quad ;\quad \eta =B^{-1/2}\varphi _{t}.
\label{new_coor}
\end{equation}%
(notice that $B^{-1/2}$ is well--defined because $B$ is positive and
invertible) and the equations of motion (\ref{KG}) for the field $\varphi $
take the simple symmetric form:
\begin{equation}
\frac{\mathrm{d} }{\mathrm{d} t}%
\begin{pmatrix}
\xi  \\
\eta
\end{pmatrix}%
=%
\begin{pmatrix}
0 & B \\
-B & 0%
\end{pmatrix}%
\begin{pmatrix}
\xi  \\
\eta
\end{pmatrix}%
.  \label{KG_matrix}
\end{equation}

Thus the equations of motion for the Klein--Gordon field constitute an
infinite dimensional extension of the dynamics of a finite number of
independent oscillators (\ref{symm_dyn}). Using the Fourier expansion of the
fields $\xi $ and $\eta $ with respect to the eigenfunctions $\Phi _{k}$ of $%
B$, $\xi (x)=\sum_{k=1}^{\infty }\xi _{k}\Phi _{k}(x),\eta
(x)=\sum_{k=1}^{\infty }\eta _{k}\Phi _{k}(x)$, then, the mechanical
variables $q_{k}=\sqrt{\omega _{k}}\xi _{k}$ and $p_{k}=\eta _{k}/\sqrt{%
\omega _{k}}$ can be interpreted as position and momentum for a
one--dimensional oscillator of frequency $\omega _{k}$ and their evolution
in time, given by eq. (\ref{standard}), as a trajectory in phase space $%
\Omega =\mathbb{R}^{2\infty }$. In the presence of field fluctuations we
have to introduce a statistical interpretation to the mechanical degrees of
freedom $(q_{k},p_{k})$ or $(\xi _{k},\eta _{k})$ of the field $\varphi (x)$%
, thus the classical statistical description of the field whose physical
meaning corresponds to the probability of a certain fluctuation of the field
to take place, will be provided by a probability law $\rho $ on the infinite
dimensional phase space $\mathbb{R}^{2\infty }$. Thus in the presence of
field fluctuations the state of the field will induce a marginal probability
density on each mode $\rho _{k}(q_{k},p_{k})$ defined by,
\begin{equation}
\rho _{k}(q_{k},p_{k})=\int \rho (q_{1},q_{2},\ldots ,q_{k},\ldots
;p_{1},p_{2},\ldots ,p_{k},\ldots )\prod_{l\neq k}\mathrm{d}q_{l}\mathrm{d}%
p_{l}.
\end{equation}%
Such marginal probability could be understood as a probability density for
the $k$--th mode of the field $\varphi $ described by the one--dimensional
oscillator with Hamiltonian $H_{k}(\xi _{k},\eta _{k})$. Similar
considerations could be applied to finite dimensional subspaces of modes of
the field whose statistical and tomographic description would be made as in
the previous section.

The canonical or Gibbs state for the field $\varphi (x)$ is given by the
probability distribution on the infinite dimensional phase space of the
system as:
\begin{equation}
\rho _{\mathrm{can}}(\xi _{1},\xi _{2},\ldots ;\eta _{1},\eta _{2},\ldots
)=N\exp \left[ {-\frac{1}{2}\beta \sum_{k\geq 1}\omega _{k}(\xi
_{k}^{2}+\eta _{k}^{2})}\right]
\end{equation}%
with the normalization constant $N$ to be determined by regularizing the
integral:
\begin{equation}
\int e^{-\frac{1}{2}\beta \sum_{k\geq 1}\omega _{k}(\xi _{k}^{2}+\eta
_{k}^{2})}\prod_{k=1}^{\infty }\mathrm{d}\xi _{k}\mathrm{d}\eta _{k}=\left[
\det \left( \frac{1}{2}\beta B\right) \right] ^{-1},
\end{equation}%
what amounts to define the determinant of the operator $B$ by using the $%
\zeta $--function regularization of determinants, i.e.,
\begin{equation}
\det \left( \frac{1}{2}\beta B\right) =\exp \left[ \zeta _{\frac{1}{2}\beta
B}^{\prime }(0)\right] ,
\end{equation}%
with
\begin{equation}
\zeta _{\frac{1}{2}\beta B}(s)=\sum_{k=1}^{\infty }\left( \frac{1}{2}\beta
\omega _{k}\right) ^{-s}.
\end{equation}%
In other words, the canonical ensemble for the real scalar Klein--Gordon
field $\varphi (x)$ is defined as the Gaussian measure with variance $C=(%
\frac{1}{2}\beta B)^{-1}$ on $\mathbb{R}^{2\infty }$. Notice that
\begin{eqnarray}
H\left[ \boldsymbol{\xi },\boldsymbol{\eta }\right]  &=&\frac{1}{2}%
\sum_{k=1}^{\infty }\omega _{k}(\xi _{k}^{2}+\eta _{k}^{2})=\frac{1}{2}%
\left\vert \left\vert B\varphi \right\vert \right\vert ^{2}+\frac{1}{2}%
\left\vert \left\vert \varphi _{t}\right\vert \right\vert ^{2}=H\left[
\varphi \right]  \\
&=&\frac{1}{2}\int_{\mathcal{V}}(\partial _{\mu }\varphi \partial ^{\mu
}\varphi +m^{2}\varphi ^{2})\mathrm{d}^{d}x,  \notag
\end{eqnarray}%
with $\frac{1}{2}||B\varphi ||^{2}$ denoting the potential $U\left[ \varphi %
\right] $ of the Klein--Gordon field in the Hamiltonian picture. Observe
that $U\left[ \varphi \right] $ can also be written as:
\begin{equation}
U\left[ \varphi \right] =\frac{1}{2}||B\varphi ||^{2}=\frac{1}{2}\langle
\varphi ,B^{2}\varphi \rangle =\frac{1}{2}\int_{\mathcal{V}}\varphi
(x)(-\Delta +m^{2})\varphi (x)\mathrm{d}^{d}x.
\end{equation}%
Then the canonical ensemble for the Klein--Gordon field at finite
temperature will be written in the usual form:
\begin{equation}
\mathrm{d}\mu _{\mathrm{can}}\left[ \varphi \right] =N\mathrm{e}^{-\frac{%
\beta }{2}\int_{\mathcal{V}}(\partial _{i}\varphi \partial ^{i}\varphi
+m^{2}\varphi ^{2})\mathrm{d}^{d}x}\mathcal{D}\varphi   \label{can_state}
\end{equation}%
with $\mathcal{D}\varphi =\prod_{k=1}^{\infty }\mathrm{d}q_{k}\mathrm{d}p_{k}
$. Moreover, if $F\left[ \varphi \right] $ denotes an observable on the
field $\varphi $ (like the energy, momentum, etc.), then the expected value
of $F$ on the canonical distribution at temperature $\beta $ will be given
by:
\begin{equation}
\langle F\rangle _{\mathrm{can}}=\frac{\int F\left[ \varphi \right] \mathrm{e%
}^{-\beta H\left[ \varphi \right] }\mathcal{D}\varphi }{\int \mathrm{e}%
^{-\beta H\left[ \varphi \right] }\mathcal{D}\varphi }.
\end{equation}%
The tomographic description of the states of the Klein--Gordon field will be
performed as in the case of an ensemble of harmonic oscillators in section %
\ref{harmonic} by choosing the space $\mathcal{M}$ the phase space $\mathbb{R%
}^{2\infty }$ itself and $\mathcal{N}=\prod_{k=1}^{\infty }\mathcal{N}_{k}$
with $\mathcal{N}_{k}$ the space of straight lines on the phase space of the
one--dimensional oscillator $(\xi _{k},\eta _{k})$. Then, as in $(\ref%
{class_tom_oscillators})$, we will define:
\begin{eqnarray}
\mathcal{W}_{\rho _{\mathrm{can}}}\left[ \boldsymbol{X},\boldsymbol{\mu },%
\boldsymbol{\nu }\right]  &=&\int \rho _{\mathrm{can}}\left[ \boldsymbol{\xi
},\boldsymbol{\eta }\right] \prod_{k=1}^{\infty }\delta (X_{k}-\mu _{k}\xi
_{k}-\nu _{k}\eta _{k})d\xi _{k}d\eta _{k}  \label{continuous_fields} \\
&=&\int e^{-\beta H\left[ \boldsymbol{\xi },\boldsymbol{\eta }\right]
}\delta \left[ X(x)-\mu (x)\xi (x)-\nu (x)\eta (x)\right] \mathcal{D}\xi \,%
\mathcal{D}\eta   \notag
\end{eqnarray}%
Here the Dirac functional distribution must be understood as an infinite
continuous product:
\begin{eqnarray}
&&\delta \left[ X(x)-\mu (x)\xi (x)-\nu (x)\eta (x)\right] =\prod_{k}\delta
\left( X_{k}-\mu _{k}\xi _{k}-\nu _{k}\eta _{k}\right)  \\
&=&\int \exp \left[ i\int K(x)\left( X(x)-\mu (x)\xi (x)-\nu (x)\eta
(x)\right) \mathrm{d}^{d}x\right] \mathcal{D}K,  \notag
\end{eqnarray}%
where $X(x)$, $\mu (x)$ and $\nu (x)$ are fields whose expansion on the
modes $\omega _{k}$ of the field $\varphi (x)$ are given respectively by:
\begin{equation}
X(x)=\sum_{k=1}^{\infty }X_{k}\Phi _{k}(x);\quad \mu (x)=\sum_{k=1}^{\infty
}\mu _{k}\Phi _{k}(x);\quad \nu (x)=\sum_{k=1}^{\infty }\nu _{k}\Phi _{k}(x).
\label{expansions}
\end{equation}%
Notice that the time dependence of the various fields is encoded in the
coefficients of the corresponding expansions. Taking
advantage again of the scaling property of the delta function we may use the
natural parametrization of optical tomograms defined by the
reparametrization $\tilde{\mu}_{k}=\mu _{k}/\sqrt{\mu _{k}^{2}+\nu _{k}^{2}}%
=\cos \theta _{k}$, $\tilde{\eta}_{k}=\nu _{k}/\sqrt{\mu _{k}^{2}+\nu
_{k}^{2}}=\sin \theta _{k}$, $\tilde{X}_{k}=X_{k}/\sqrt{\mu _{k}^{2}+\nu
_{k}^{2}}$ and after standard computations we get:
\begin{equation}
\mathcal{W}_{\rho _{\mathrm{can}}}^{\mathrm{opt}}(\boldsymbol{\tilde{X}},%
\boldsymbol{\theta })=N\mathrm{e}^{-\sum_{k=1}^{\infty }\tilde{X}_{k}^{2}}=N%
\mathrm{e}^{-\int_{\mathcal{V}}\tilde{X}(x)^{2}\mathrm{d}^{d}x}=N\mathrm{e}%
^{-||\boldsymbol{\tilde{X}}||^{2}}.
\end{equation}%
with $\theta (x)=\tan ^{-1}\left[ \eta (x)/\xi (x)\right] $ and the
normalization constant $N$ defined by choosing a proper regularization of
the trace of the operator $B$.

\section{Tomographic picture of continuous modes}

If we consider the scalar field in an infinite volume cavity or in the full
Minkowski space--time for instance, many or all of the modes of the system
will become continuous. For simplicity we will assume that we are discussing
the field in the $d+1$ Minkowski space--time $\mathbb{R}^{d+1}$ and the
continuous modes of the fields $\varphi (x)$, $\xi (x)$, $\eta (x)$ are
described by the wave vector $k$, say,
\begin{equation}
\xi (x)=\frac{1}{(2\pi )^{d/2}}\int \left( \xi _{k}\mathrm{e}^{-\mathrm{i}%
k\cdot x}+\xi _{-k}\mathrm{e}^{\mathrm{i}k\cdot x}\right) \mathrm{d}^{d}k,
\end{equation}%
etc. Now a state of the field $\varphi (x)$ will be represented by a
probability measure $\rho \left[ \xi ,\eta \right] $, again nonnegative and
normalized. An example of such state will be given by the canonical
ensemble, this is the Gaussian measure whose covariance is the operator $B$
as in (\ref{can_state}):
\begin{equation}
d\mu _{\mathrm{can}}\left[ \varphi \right] =\mathrm{e}^{-\beta H\left[
\varphi \right] }\mathcal{D}\varphi =\mathrm{e}^{-\beta H\left[ \xi ,\eta %
\right] }\mathcal{D}\xi \mathcal{D}\eta
\end{equation}%
with the normalization constant absorbed in the definition of the measure.

We will consider as analogue of Gibbs states, states that are absolutely
continuous with respect to the canonical state, i.e., states of the form:
\begin{equation}
\rho \left[ \varphi \right] =f\left[ \xi (x),\eta (x)\right] \mu _{\mathrm{%
can}}  \label{statesfield}
\end{equation}%
with
\begin{eqnarray}
&&f\left[ \xi (x),\eta (x)\right] \geq 0 \\
&&\int f\left[ \xi (x),\eta (x)\right] e^{-\beta H\left[ \xi ,\eta \right] }%
\mathcal{D}\xi \mathcal{D}\eta =1.
\end{eqnarray}

Even though at a formal level, we may introduce as in (\ref%
{continuous_fields}) a tomographic probability density for a state of the
field of the form (\ref{statesfield}) as a functional of three auxiliary
tomographic fields $X(x),\xi (x),\eta (x)$ and apply, at the functional
level, the usual Radon transform. The expansions (\ref{expansions}) will be
replaced by the Fourier transform:
\begin{equation}
X(x)=\frac{1}{(2\pi )^{d/2}}\int \left( X_{k}\mathrm{e}^{-\mathrm{i}k\cdot
x}+X_{-k}\mathrm{e}^{\mathrm{i}k\cdot x}\right) \mathrm{d}^{d}k,\quad
\mathrm{etc.}
\end{equation}%
Then,
\begin{equation}
\mathcal{W}_{f}\left[ X(x),\mu (x),\nu (x)\right] =\int f\left[ \xi (x),\eta
(x)\right] \delta \left[ X(x)-\mu (x)\xi (x)-\nu (x)\eta (x)\right]
e^{-\beta H[\xi ,\eta ]}\mathcal{D}\xi \mathcal{D}\eta .  \label{newWfield}
\end{equation}%
The inverse Radon transform maps the tomographic probability density given
by (\ref{newWfield}) onto the probability density functional
\begin{equation}
\rho _{f}\left[ \xi ,\eta \right] =\int \mathcal{W}_{f}\left[ X,\mu ,\nu %
\right] \exp \left[ \mathrm{i}(X(x)-\mu (x)\xi (x)-\nu (x)\eta (x))\right]
\mathcal{D}X(x)\mathcal{D}\mu (x)\mathcal{D}\nu (x)  \label{inverse_funct}
\end{equation}%
The tomographic probability functional (\ref{newWfield}) has the properties
of nonnegativity and normalization, i.e.
\begin{eqnarray}
&&\mathcal{W}_{f}\left[ X(x),\mu (x),\nu (x)\right] \geq 0 \\
&&\int \mathcal{W}_{f}\left[ X(x),\mu (x),\nu (x)\right] \mathcal{D}X(x)=1.
\end{eqnarray}%
These formulas hold true for any value of the auxiliary fields $X(x),\mu
(x),\nu (x)$.

In the current case the manifold $\mathcal{N}$ used to construct the
generalized Radon transform is described by the tomographic fields $X(x),\nu
(x),\mu (x)$, which would be a continuum version of the finite--mode version
of the straight lines:
\begin{equation}
X_{k}-\mu _{k}\xi _{k}-\nu _{k}\eta _{k}=0.
\end{equation}

We will end this discussion by emphasizing again the homogeneity property of
the tomographic description of the scalar field we just presented,
homogeneity that is described by the condition:
\begin{equation}
\left[ X(x)\frac{\delta }{\delta X(x)}+\mu (x)\frac{\delta }{\delta \mu (x)}%
+\nu (x)\frac{\delta }{\delta \nu (x)}+1\right] \mathcal{W}_{f}\left[
X(x),\mu (x),\nu (x)\right] =0
\end{equation}

\section{The tomographic picture of evolution equation for classical fields}

In the previous sections we have seen that the state of the classical scalar field $\varphi (x)$ can be described
either by a probability density functional $f\left[ \xi (x),\eta (x)\right] $
on the field phase--space or by the tomographic probability density
functional $\mathcal{W}_{f}\left[ X(x),\mu (x),\nu (x)\right] $. Both
probability density functionals are connected by the invertible functional
Radon transform (\ref{newWfield}), (\ref{inverse_funct}) and in view of
this, they both contain equivalent information on the random field states.
The dynamical evolution of states of the field $\varphi (t,x)$ will be
determined by the Klein--Gordon equation (\ref{KG_matrix})

If the Hamiltonian providing the evolution of the field is given by the sum
of kinetic and potential energy
\begin{equation}
\mathcal{H}\left[ \varphi \right] =\frac{1}{2}\int \left( \dot{\varphi}%
(x)^{2}+V\left[ \varphi (x)\right] \right) \mathrm{d}^{d}x=\frac{1}{2}||\dot{%
\varphi}||^{2}+U\left[ \varphi \right] ,  \label{Ham}
\end{equation}%
the evolution of the probability density functional on the classical
phase--space of the field obeys a Liouville functional differential
equation:
\begin{equation}
\frac{\mathrm{d} f}{\mathrm{d} t}=\{f,\mathcal{H}\}.  \label{Lioufree}
\end{equation}%
The functional Poisson brackets above are given by:
\begin{equation}
\{F\left[ \xi ,\eta \right] ,G\left[ \xi ,\eta \right] \}=\int \left( \frac{%
\delta F}{\delta \xi (x)}\left\{ {\xi (x),\eta (y)}\right\} \frac{\delta G}{%
\delta \eta (y)}+\frac{\delta F}{\delta \eta (x)}\left\{ {\eta (x),\xi (y)}%
\right\} \frac{\delta G}{\delta \xi (y)}\right) \mathrm{d}^{d}x\mathrm{d}%
^{d}y,
\end{equation}%
where the fields $\xi (x)$, $\eta (y)$ satisfy the relations:
\begin{equation}
\{\xi (x),\eta (y)\}=\delta ^{d}(x-y).
\end{equation}%
Then we obtain for the Hamiltonian $\mathcal{H}$ above (\ref{Ham}) the
expression:
\begin{equation}
\frac{\mathrm{d}}{\mathrm{d} t}f\left[ \xi ,\eta \right] +\int \eta (x)\frac{%
\delta f\left[ \xi ,\eta \right] }{\delta \xi (x)}\mathrm{d}^{d}x-\int \frac{%
\delta V\left[ \xi ,\eta \right] }{\delta \xi (x)}\frac{\delta f\left[ \xi
,\eta \right] }{\delta \eta (x)}\mathrm{d}^{d}x=0  \label{Liou}
\end{equation}%
which is just the $n\rightarrow \infty $ limit of the Liouville equation for
finite number of field modes discussed in section \ref{harmonic}.

In the case that $V\left[ \varphi \right] =0$ we have the functional
Liouville equation
\begin{equation}
\frac{\mathrm{d} }{\mathrm{d} t}f\left[ \xi ,\eta \right] +\int \eta (x)\frac{%
\delta f\left[ \xi ,\eta \right] }{\delta \xi (x)}\mathrm{d}^{d}x=0.
\end{equation}%
and the corresponding tomographic form of this equation reads
\begin{equation}
\frac{\mathrm{d} }{\mathrm{d} t} \mathcal{W}_{f}\left[ X,\mu ,\nu \right]-\int
\mu (x)\frac{\delta \mathcal{W}_{f}\left[ X,\mu ,\nu \right] }{\delta \nu (x)%
}\mathrm{d}^{d}x=0.
\end{equation}

To get this equation starting from (\ref{Lioufree}) we used the
correspondences:
\begin{eqnarray}
&&\frac{\delta }{\delta \xi (x)}\leftrightarrow \mu (x)\frac{\delta }{\delta
X(x)};\frac{\delta }{\delta \eta (x)}\leftrightarrow \nu (x)\frac{\delta }{%
\delta X(x)}  \notag  \label{map} \\
&&\xi (x)\leftrightarrow -\frac{\delta }{\delta \mu (x)}\left[ \frac{\delta
}{\delta X(x)}\right] ^{-1};\eta (x)\leftrightarrow -\frac{\delta }{\delta
\nu (x)}\left[ \frac{\delta }{\delta X(x)}\right] ^{-1}.
\end{eqnarray}

These relations correspond to a realization of the infinite Heisenberg--Weyl
algebra generators (and enveloping algebra) on the field phase--space and
the map of the representation in terms of the generator action onto the
tomograms.

The rule (\ref{map}) provide a possibility to construct the tomographic form
of the Liouville equation (\ref{Liou}). Using the substitution ${%
f\rightarrow \mathcal{W}_{f}}$ in (\ref{Liou}), and the substitutions (\ref%
{map}), we get the field evolution equation
\begin{eqnarray}
&&\frac{\mathrm{d} }{\mathrm{d} t} \mathcal{W}_{f}\left[ X,\mu ,\nu \right] =\int
\,\mathrm{d}^{d}x^{\prime }\mu (x^{\prime })\frac{\delta \mathcal{W}_{f}%
\left[ X(x),\mu (x),\nu (x)\right] }{\delta \nu (x^{\prime })}  \notag
\label{evfield} \\
&&+\int \mathrm{d}^{d}x^{\prime }\left[ \frac{\delta V}{\delta \xi
(x^{\prime })}\left( \xi \left( x\right) \rightarrow -\frac{\delta }{\delta
\mu (x)}\left[ \frac{\delta }{\delta X(x)}\right] ^{-1}\right) \nu
(x^{\prime })\frac{\delta }{\delta X(x^{\prime })}\right] \mathcal{W}_{f}%
\left[ X,\mu ,\nu \right] .
\end{eqnarray}

For the case of field which is a collection of noninteracting oscillators
described by the potential energy
\begin{equation}
V\left[ \varphi \right] =\frac{1}{2}m^{2}\varphi ^{2}  \label{Energy}
\end{equation}%
then (\ref{evfield}) reads
\begin{equation}
\frac{\mathrm{d} }{\mathrm{d} t} \mathcal{W}_{f}=\int \mathrm{d}^{d}x\left( \mu
(x)\frac{\delta \mathcal{W}_{f}}{\delta \nu (x)}-\nu (x)\frac{\delta
\mathcal{W}_{f}}{\delta \mu (x)}\right) .  \label{evfieldenergy}
\end{equation}%
which is the equivalent of eq. (\ref{evol_tomo_class}) to the continuous
scalar field $\varphi $.

\section{Conclusion and perspectives}

A proposal for the tomographic description of a family of statistical states
for a classical real scalar Klein-Gordon field has been presented inspired
by the tomographic description of statistical states for an ensemble of
harmonic oscillators. This tomographic description of classical fields
shares most of the tomographic properties of tomograms for classical states:
homogeneity, positivity and normalization. Moreover the field equations,
represented as the evolution equation for field states, are reproduced in
tomographic terms, paving the way towards a tomographic description of the
quantum scalar field. Notice that the tomographic description presented in
this work, a natural extension of Radon transform, breaks the Lorentz
covariance of the field theory, thus the Lorentz covariance of the
tomographic description should be restored at the end. Lorentz covariance,
as well as gauge invariance (when interactions are introduced), should be
incorporated as a natural ingredient in the tomographic picture. The
tomographic picture of other fields like Maxwell, Dirac, Proca, Einstein
could be addressed following similar arguments. Such issues will be
discussed in subsequent works.


\begin{thebibliography}{99}
\bibitem{Sc26} E. Schr\"{o}dinger, Annalen d. Physik \textbf{79} (1926) 361
; \textbf{81} (1926) 109.

\bibitem{He27} W. Heisenberg, Z. Phys. \textbf{43} (1927) 172.

\bibitem{Wi32} E. Wigner, Phys. Rev. \textbf{40} (1932) 749.

\bibitem{Pedatom} A. Ibort, V. I. Man'ko, G. Marmo, A. Simoni, and F.
Ventriglia, Phys. Scr. \textbf{79} (2009) 065013.

\bibitem{VentriPositive} A. Ibort, V. I. Man'ko, G. Marmo, A. Simoni, and F.
Ventriglia, Phys. Lett. A \textbf{374} (2010) 2614.

\bibitem{JPA2002} O. V. Man'ko, V. I. Man'ko, and G. Marmo, J. Phys. A:
Math. Gen. \textbf{35} (2002) 699.

\bibitem{Ra17} J. Radon. \emph{\"{U}ber die bestimmung von funktionen durch
ihre integralwerte l\"{a}ngs gewisser mannigfaltigkeiten}. Ber. Ver. S\"{a}%
chs. Akad. Wiss. keipzing. mahtj-Phys., KL., 69:262--277 (1917). English
translation in S.R. Deans: The Radon Transform and Some of Its applications,
app. A Krieger Publ. Co., Florida 2nd ed. (1993).

\bibitem{Manko-Tombesi} O. V. Man'ko, V. I. Man'ko, J. Russ. Laser Res.
\textbf{18} (1997) 407.

\bibitem{Me00} V. I. Man'ko and R. Vilela Mendes, Physica D \textbf{145}
(2000) 330-348.

\bibitem{Ma98} V. I. Man'ko, L. Rosa, and P. Vitale, Phys. Lett. B \textbf{%
439} (1998) 328.

\bibitem{St05} V. I. Man'ko, G. Marmo and C. Stornaiolo, Gen. Relativ.
Gravit. \textbf{37} (2005) 99; ibid., Gen. Relativ. Gravit. \textbf{37}
(2005) 2003.

\bibitem{An09} V.A. Andreev, M.A. Man'ko, V.I. Man'ko, N.C. Thanh, N.H. Son,
S.D. Zakharov. J. Russ. Laser Res. \textbf{30} (2009) 591-598.

\bibitem{Ma09} M. A. Man'ko, V. I. Man'ko, N. C. Thanh, N. H. Son, Y. P.
Timoreev, S. D. Zakharov. J. Russ. Laser Res. \textbf{30} (2009) 1-11.

\bibitem{asorey} M. Asorey, P. Facchi, V. I. Man'ko, G. Marmo, S. Pascazio, E. C. G. Sudarshan,
 Phys. Rev. A \textbf{76} (2007) 012117.

\bibitem{Ib11} A. Ibort, V. I. Man'ko, G. Marmo, A. Simoni and F.
Ventriglia. Phys. Scr. \textbf{84} (2011) 065006.

\bibitem{sudklaudbook} J. Klauder and E. C. G. Sudarshan, \textit{Fundamentals of Quantum Optics}, Benjamin (New York, 1968), (reprint: Dover, 2006).

\bibitem{dodman}  V. V. Dodonov, V. I. Manko, \textit{Invariants and Evolution of Nonstationary Quantum Systems. Proceedings of the P.N. Lebedev Physical Institute}, Vol. 183 (ed. M. A. Markov), Nova Science (New York 1989).

\bibitem{GraRyz} I. S. Gradshteyn, I. M. Ryzhik, \textit{Table of Integrals,
Series, and Products. Seventh Edition.} Elsevier-Academic Press (USA 2007).

\bibitem{Ch07} V.N. Chernega, V.I. Man'ko. J. Russ. Laser Res. \textbf{28}
(2007) 535--547.

\bibitem{masasivibook} G. Marmo, E. J. Saletan, A. Simoni, B. Vitale, \textit{Dynamical Systems. Differential Geometrical Approach to Symmetry and Reduction}, J. Wiley  (Chichester, 1985).
\end{thebibliography}
\end{document}